\documentclass[aps,prx,twocolumn,showpacs,floatfix,superscriptaddress,graphics,longbibliography]{revtex4-1}

\usepackage{amsmath,amssymb,graphicx,color,braket,hyphenat,makeidx,xcolor,subfigure, dsfont}
\usepackage{bm}
\usepackage[retainorgcmds]{IEEEtrantools}
\usepackage{graphicx}
\usepackage{mathrsfs}
\usepackage{amsmath}
\usepackage{amssymb}
\usepackage{color}
\usepackage{amsfonts}
\usepackage{times,txfonts}
\usepackage{nicefrac}

\AtBeginDocument{%
    \newwrite\bibnotes
    \def\bibnotesext{Notes.bib}
    \immediate\openout\bibnotes=\jobname\bibnotesext
    \immediate\write\bibnotes{@CONTROL{REVTEX41Control}}
    \immediate\write\bibnotes{@CONTROL{%
    apsrev41Control,author="08",editor="1",pages="1",title="0",year="1"}}
     \if@filesw
     \immediate\write\@auxout{\string\citation{apsrev41Control}}%
    \fi
}%

\usepackage[colorlinks=true,linkcolor=blue,urlcolor=blue,citecolor=red,pdfusetitle]{hyperref}

\newcommand{\covy}{\mathrm{cov}_y}
\newcommand{\var}{\mathrm{var}}
\newcommand{\tr}{\mathrm{Tr}}

\newcommand{\id}{{\mathds 1}}
\newcommand{\ii}{^{(i)}}
\newcommand{\Lm}{{\mathds L}}
\newcommand{\A}{{\mathds A}}

\begin{document}

\title{Non-Abelian Quantum Transport and Thermosqueezing Effects}
\author{Gonzalo Manzano}
\affiliation{Institute for Cross-Disciplinary Physics and Complex Systems (IFISC) UIB-CSIC, Campus Universitat Illes Balears, E-07122 Palma de Mallorca, Spain}
\affiliation{Institute for Quantum Optics and Quantum Information (IQOQI), Austrian Academy of Sciences, Boltzmanngasse 3, 1090 Vienna, Austria.}
\author{Juan M. R. Parrondo}
\affiliation{Departamento de Estructura de la Materia, F\'isica T\'ermica y Electr\'onica and GISC, Universidad Complutense Madrid, E-28040 Madrid, Spain}
\author{Gabriel T. Landi}
\affiliation{Instituto de F\'isica da Universidade de S\~ao Paulo,  05314-970 S\~ao Paulo, Brazil.}

\begin{abstract}
Modern quantum experiments provide examples of transport with non-commuting quantities, offering a tool to understand the interplay between thermal and quantum effects. Here we set forth a theory for non-Abelian transport in the linear response regime.
Our key insight is to use generalized Gibbs ensembles with non-commuting charges as the basic building blocks and strict charge-preserving unitaries in a collisional setup. The linear response framework is then built using a collisional model between two reservoirs.
We show that the transport coefficients obey Onsager reciprocity.
{Moreover, we find that quantum coherence, associated to the non-commutativity, acts so as to reduce the net entropy production, when compared to the case of commuting transport.
This therefore provides a clear connection between quantum coherent transport and dissipation.} 
As an example, we study heat and squeezing fluxes in bosonic systems, characterizing a set of thermosqueezing coefficients with potential applications in metrology and heat-to-work conversion in the quantum regime.
\end{abstract}

\maketitle 

\section{Introduction}

In the simplest scenario, putting in contact two thermal systems at different temperatures causes them to exchange heat~\cite{Callen1985}. The amount of heat exchanged is proportional to the temperature gradient, which is known as Fourier's law. Similarly, a gradient of chemical concentration (or electric voltage) generates a flow of particles (or electrons), as predicted by Fick's (Ohm's) law. 
There are also cross effects: A gradient of temperature causes a flow of particles (the Seebeck effect), while a gradient of chemical potential causes a flow of heat (Peltier effect). 
These phenomena form the basis of a great variety of applications in thermoelectricity, thermomagnetism, and galvanomagnetic phenomena~\cite{Sommerfeld1931,Beretta2019}.
One of the first major advances in the construction of modern non-equilibrium thermodynamics was the development of a solid theoretical basis for explaining them, as provided by Onsager's reciprocity theory~\cite{Onsager1931, Onsager1931a}.

Heat and particle transport deals with observables which commute at the quantum level. This is what we shall refer to as Abelian transport and is known to crucially impact the thermodynamics of both classical~\cite{Horowitz2016,Rao2018} and quantum systems~\cite{Vaccaro2011,Hickey2014,Perarnau-Llobet2016,Guryanova2016a,YungerHalpern2016c,Mur-Petit2018}. \emph{Non-Abelian transport}, on the other hand, is associated with observables which do not commute, owing to the appearance of other types of quantum excitations~\cite{Halpern2018c}. Examples include the transport of energy and magnetization in certain types of spin chains~\cite{YungerHalpern2020}, such as the transverse-field Ising model~\cite{Calabrese2005a,Bertini2021};  transport of energy and particles in the Kitaev model of superconducting wires~\cite{Bhat2020}; and transport of energy and bosonic squeezing in quantum optical systems~\cite{Manzano2018b,Manzano2016}.

The fundamental implications of non-commuting conserved quantities on the thermodynamics of quantum systems has only started to be explored~\cite{Guryanova2016a,YungerHalpern2016b,Lostaglio2017b,Croucher2017,Ikeda2020} and several questions remain open.
Filling  this gap would be both valuable and timely, for at least two reasons. 
First, the phenomenon is already well within reach of several modern quantum experimental platforms, such as ultra-cold atoms~\cite{Bloch2008,Brantut2013} or optomechanical devices~\cite{Brunelli2016a,Mitchison2018}. Second, non-Abelian transport 
provides a perfect arena for understanding the interplay between thermal phenomena and quantum coherence, a critical issue which has been at the center of many quantum thermodynamic studies in the last years~\cite{Bradner2017,Kwon2018b,Holmes2018a,Henao2018,Santos2019,Francica2019,Manzano2019,Lostaglio2019,Rodrigues2019,Latune2019,Micadei2019}. Since exchanges of extra conserved quantities may contribute to both heat and work, a non-Abelian transport theory may help to clarify the thermodynamics of squeezing-powered engines~\cite{Ronagel2014,Klaers2017a,Manzano2017b,Klaers2019} and other 
non-thermal devices~\cite{Scully2003,Abah2014,Niedenzu2018}.

\begin{figure}[t]
\includegraphics[width= 1.0 \linewidth]{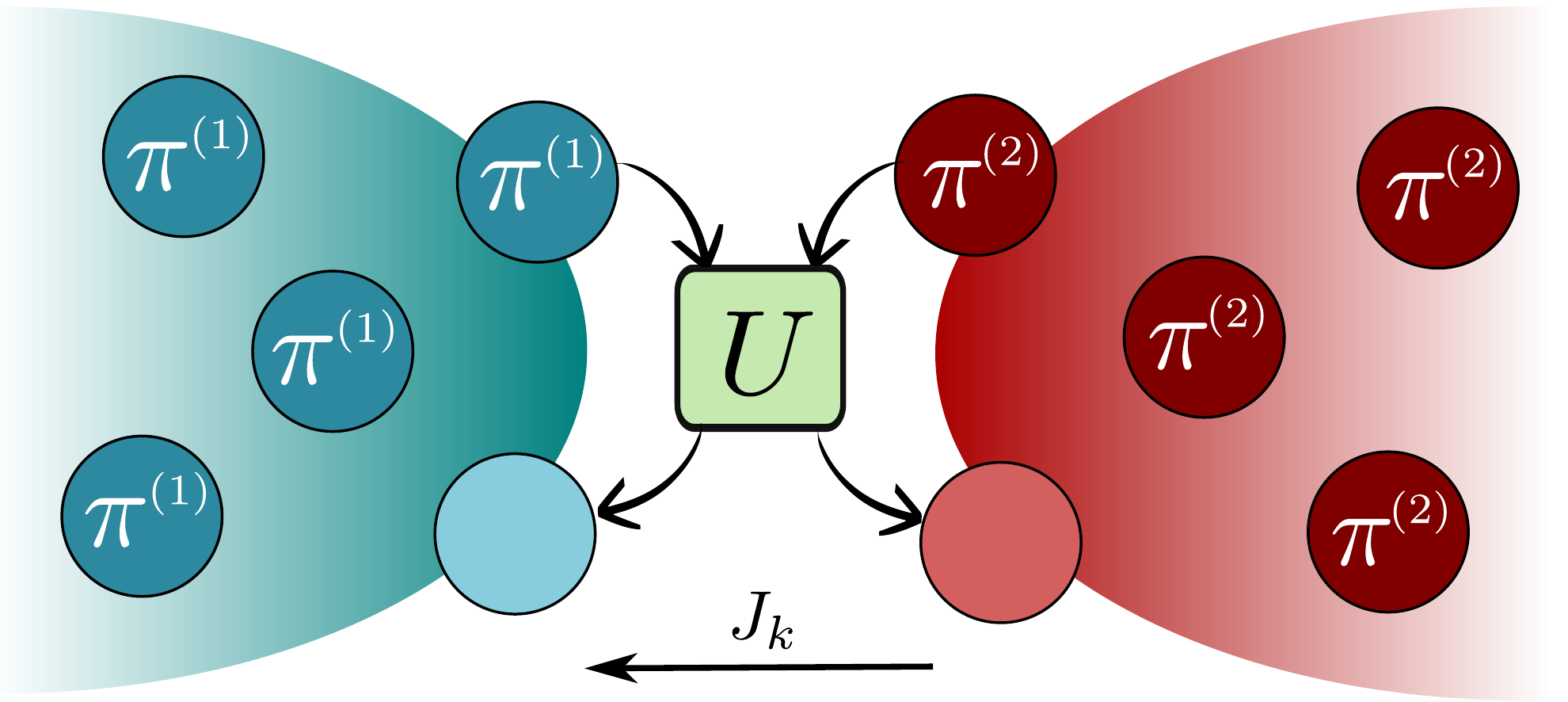}
\caption{Collisional transport framework between two reservoirs (red and blue regions). At each time step two subsystems prepared in GGE states $\pi^{(1)}$ and $\pi^{(2)}$ interact through a charge-preserving unitary operation $U$ (green box). After interacting, the two subsystems are returned to their respective reservoirs, producing a net flow of charges $J_k$.} \label{fig:1}
\end{figure}

In this paper we establish a framework for non-Abelian transport in the linear response regime,  generalizing Onsager's reciprocity theory to the case of arbitrary non-commuting charges. 
Non-commutativity breaks joint fluctuation theorems, which are often used to derive Onsager reciprocity~\cite{Esposito2009PRL}. It also leads to a disconnection between different notions of equilibrium, associated to the maximum entropy principle and complete passivity~\cite{YungerHalpern2016b,Lostaglio2017b}.
To overcome these difficulties, we introduce a collisional model, which allows us to formulate non-Abelian transport in terms of the notion of charge preservation. 

Our results shows that the transport coefficients, as well as the entropy production, can be written in terms of the so-called y-covariances~\cite{Kubo1985,Petz2002,Miller2019,Scandi2020,Miller2021}, which provide a generalization of the notion of covariances to non-commuting observables. 
This is further expressed in terms of the Wigner-Yanase-Dyson skew information~\cite{Petz2002}, a quantifier of quantum coherence. Inspired by~\cite{Miller2019,Scandi2020}, we then use this to pinpoint the role of coherence in the resulting transport properties. 
{More specifically, we find that quantum coherence, associated to the non-commutativity of the thermodynamic charges, acts so as to \emph{reduce} the total entropy production of the process.}
As an application, we study the transport of heat and squeezing in bosonic systems. 
Our framework allows the introduction and characterization of the squeezing version of the Seebeck and Peltier effects. The interesting avenues of research this opens are exemplified by metrology in a squeezing-based thermocouple, and heat-to-work conversion in an autonomous engine.

\section{Formal Framework} \label{sec:framework}
Let $\{Q_k\}$ denote a set of not necessarily commuting observables, which we henceforth refer to as thermodynamic \emph{charges} (e.g. the Hamiltonian $H$ and number of particles operators $N$ in the Abelian case). We assume each charge is associated with a corresponding affinity $\lambda_k$, as described by the Generalized Gibbs Ensemble (GGE)~\cite{Jaynes1957}
\begin{equation} \label{GGE}
\pi_{\bm{\lambda}}
= \frac{e^{-\sum_k \lambda_k Q_k}}{Z},
\end{equation}
where $Z = \tr[e^{-\sum_k \lambda_k Q_k}]$ is the partition function.
GGEs can be viewed as the natural generalization of the grand-canonical ensemble, $e^{-\beta (H - \mu N)}/Z$, where $\beta=1/T$ is the inverse temperature and $\mu$ the chemical potential (the affinities of $H$ and $N$ are thus $1/T$ and $\mu/T$).
These states are known to play a fundamental role in describing the relaxation of integrable many-body quantum systems after a sudden quench~\cite{Langen2015,Rigol2007}.
 
We consider non-Abelian transport between two reservoirs, 1 and 2, each described by their own set of charges $\{Q_k\ii\}$ and prepared in a GGE of the form~\eqref{GGE}, but with possibly different affinities  $\bm{\lambda}^{(1)} = (\lambda_1^{(1)}, \lambda_2^{(1)}, ...)$ and $\bm{\lambda}^{(2)} = (\lambda_1^{(2)}, \lambda_1^{(2)}, ...)$. 
Each reservoir consists of an infinite pool of simple subsystems or units. To model the interaction between the two reservoirs we use a collisional model approach, that simplifies the problem and has demonstrated to be very useful in the theory of open quantum systems~\cite{Rau1963,Giovannetti2012,Ciccarello2017,DeChiara2018,Cattaneo2021}. In its most basic version, a small system interacts sequentially (or ``collides'') with fresh units coming from a reservoir. Even though the evolution in each interaction is unitary, tracing the reservoir units after interaction leads to 
dissipative dynamics on the system ruled (under mild conditions) by the same Markovian master equation as 
in typical weak-coupling perturbation scenarios~\cite{Breuer2007}. As in Ref.~\cite{DeChiara2018}, here we extend this collisional approach to the problem of transport, that is, the two reservoirs interact with each other through an infinite sequence of small and equivalent collisions as scketched in  Fig.~\ref{fig:1}.

In each collision two arbitrary subsystems, 1 and 2, in states $\pi_{\bm{\lambda}^{(1)}}^{(1)}$ and  $\pi_{\bm{\lambda}^{(2)}}^{(2)}$ interact unitarily according to the map
\begin{equation}\label{global_map}
\rho_{12}' = U\big(\pi_{\bm{\lambda}^{(1)}}^{(1)} \otimes \pi_{\bm{\lambda}^{(2)}}^{(2)}\big) U^\dagger, 
\end{equation} 
where $U$ is a unitary operator~\cite{noteH}.
In order to enforce transport during the interaction, we assume that $U$ is charge preserving; that is, it satisfies~\cite{Brandao2013,Horodecki2013,Brandao2015}
\begin{equation}\label{charge_preservation}
[U, Q_k^{(1)} + Q_k^{(2)}] = 0,\qquad \forall k.
\end{equation}
This implies that any increase of a charge in subsystem 1, $Q_k^{(1)}$, will always be associated to an equal decrease of the same charge in subsystem 2, $Q_k^{(2)}$.
This property is crucial in order to have proper \emph{transport}; 
it would make no sense to talk about a quantity being ``transported'', if it could be spontaneously created or destroyed.   
When Eq.~\eqref{charge_preservation} is satisfied, one may  unambiguously define the current associated to the charge $Q_k$ per collision as 
\begin{eqnarray}
J_k := \langle Q_k^{(1)} \rangle_\mathrm{f} - \langle Q_k^{(1)}  \rangle_\mathrm{i} = -  \langle Q_k^{(2)} \rangle_\mathrm{f}+ \langle Q_k^{(2)}  \rangle_\mathrm{i},
\end{eqnarray}
where $\langle \ldots \rangle_\mathrm{f}$ and $\langle \ldots \rangle_\mathrm{i}$ refer to expectation values in the final and initial states of the map~\eqref{global_map} respectively. 
Equation~\eqref{charge_preservation} implies that the fixed point of~\eqref{global_map} is always the global equilibrium state $\pi := \pi_{\bm{\lambda}}^{(1)} \otimes \pi_{\bm{\lambda}}^{(2)}$,  where both systems have the same affinities. As a consequence, the currents will only flow provided there is a gradient of affinities.

The net entropy produced per collision can then be written in the usual form, as a product of fluxes times forces as~\cite{Callen1985,Perarnau-Llobet2016}  
\begin{equation}\label{Pi_gen}
{\Sigma} = \sum\limits_k \delta \lambda_k J_k \geqslant 0, 
\end{equation}
where $\delta \lambda_k = \lambda_k^{(1)} - \lambda_k^{(2)}$. The positivity of ${\Sigma}$ can be made manifest by expressing it in terms of information theoretic quantities~\cite{Manzano2017a,Santos2019,Strasberg2016,Esposito2010a} (see appendix \ref{sm:1} for details).

\section{Onsager coefficients} \label{sec:theorem} 
We are interested in the linear response regime, which follows by setting $\lambda_k^{(2)} = \lambda_k$ and $\lambda_k^{(1)} = \lambda_k + \delta \lambda_k$, where $\delta \lambda_k$ is assumed to be small. In the standard paradigm, the currents $J_k$ respond linearly to the affinity differences $\delta \lambda_k$ to leading order, namely: 
\begin{equation}\label{LR}
J_k = \sum\limits_\ell L_{k\ell} \delta \lambda_\ell,
\end{equation}
where $L_{k\ell}$ represent the transport coefficients of the model and form the so-called \emph{Onsager matrix} $\Lm$~\cite{Onsager1931}.
The entropy production in Eq.~\eqref{Pi_gen} then becomes a quadratic form, 
\begin{eqnarray}
\Sigma {= \sum_{k, l} L_{k,l} \delta \lambda_k \delta \lambda_l} = \bm{\delta \lambda}~\Lm~\bm{\delta \lambda}^T
\end{eqnarray}
with $\bm{\delta \lambda} = (\delta \lambda_1, \delta \lambda_2, \dots)$.
The exact form of  $\Lm$ can be found from the underlying dynamics.
The fact that the charges are non-commuting, however, introduces fundamental modifications to the usual linear response treatments. 
Notwithstanding, it turns out that $L_{k\ell}$ can be expressed in a convenient, and particularly illuminating form, as summarized in the following theorem.

\noindent
{\bf \emph{ Theorem~I}} ---\emph{Provided the set of charges $\{Q_k\}$, as well as the dynamics, are time-reversal invariant, the Onsager coefficients $L_{k\ell}$ associated to the map~\eqref{global_map} can be written as}
\begin{equation}\label{onsager}
	L_{k\ell} = \frac{1}{2} \int\limits_0^1 dy \; \covy\Big(\tilde{Q}_k^{(1)} - Q_k^{(1)}, \tilde{Q}_\ell^{(1)}  - Q_\ell^{(1)}\Big),
\end{equation}
\emph{where} $\tilde{Q}_k^{(1)} := U^\dagger Q_k^{(1)} U$ \emph{and} 
\begin{equation}\label{covy}
	\covy(A,B) = \tr\big(A \pi^y B \pi^{1-y}\big) - \tr(A\pi) \tr(B\pi), 
\end{equation}
\emph{is the so-called }``y-covariance''~\cite{Kubo1985,Petz2002} \emph{evaluated in the global equilibrium state} $\pi = \pi_{\bm{\lambda}}^{(1)} \otimes \pi_{\bm{\lambda}}^{(2)}$.
\\

The proof of the theorem can be found in appendix \ref{sm:2}. It follows from the expansion of the initial state of system 1, to leading order in $\delta \lambda_k$, as
\[
\pi_{\bm{\lambda}^{(1)}}^{(1)} \simeq \pi_{\bm{\lambda}}^{(1)} + \sum \limits_\ell \delta \lambda_\ell \frac{\partial \pi_{\bm{\lambda} + \delta \bm{\lambda}}^{(1)}}{\partial (\delta \lambda_\ell)} \bigg|_{\delta\lambda_\ell = 0} .
\]
Plugging this expansion into the currents expressions $J_k = \langle Q_k^{(1)} \rangle_\mathrm{f} - \langle Q_k^{(1)}  \rangle_\mathrm{i}$, and using the global map~\eqref{global_map}, allows us to identify the Onsager coefficients as 
\begin{equation}\label{SM_L_original}
L_{k\ell} = \tr[ (\tilde{Q}_k^{(1)} - Q_k^{(1)}) \left(\frac{\partial \pi_{\bm{\lambda} + \delta \bm{\lambda}}^{(1)}}{\partial (\delta \lambda_\ell)} \bigg|_{\delta\lambda_\ell = 0} \otimes \pi_{\bm{\lambda}}^{(2)}\right)],
\end{equation}
where $\tilde{Q}_k^{(1)} \equiv U^\dagger Q_k^{(1)} U$. Rewriting this expression by using Feynman integral representations we are able to obtain two equivalent expressions for the Onsager coefficients $L_{k \ell}$ from which, by assuming time-reversal invariant currents, the form in Eq.~\eqref{onsager} is obtained.
The case of non time-reversal invariant systems is treated in appendix~\ref{app:O-C}.

The y-covariance in Eq.~\eqref{covy}, often used in condensed matter~\cite{Kubo1985}, provides a generalization of the notion of covariance to non-commuting observables. In fact, if the process is Abelian, Eq.~\eqref{onsager} reduces to $L_{k\ell} = \frac{1}{2} \text{cov}\Big(\tilde{Q}_k^{(1)} - Q_k^{(1)}, \tilde{Q}_\ell^{(1)}  - Q_\ell^{(1)}\Big)$. But in the non-Abelian case, no such simplification exists.
\\

\noindent
{\bf \emph{Corollary~I}} --- \emph{Using the expression for the Onsager coefficients in Eq.~\eqref{onsager} the entropy production rate per collision in Eq.~\eqref{Pi_gen} can  be written as}
\begin{equation}\label{Sigma_D}
{\Sigma} = \frac{1}{2} \int\limits_0^1 dy \; \covy(D,D), 
\end{equation}
\emph{where we defined the operator $D := \sum_k \delta \lambda_k \big(\tilde{Q}_k^{(1)} - Q_k^{(1)}\big)$.}
\\

Let us now explore the main consequences of Theorem I and Corollary I. First, noticing that $\covy(A,A)\geqslant 0$ for any operator $A$~\cite{Petz2002}, we recover the positivity of the entropy production in the form~\eqref{Sigma_D}. Since ${\Sigma} \geqslant 0$ for any set of gradients $\{\delta \lambda_k\}$, it then follows that $\Lm$ is a positive semi-definite matrix. 
Second, and more importantly, using the fact that $\covy(A,B) = \text{cov}_{1-y}(B,A)$, it follows that the Onsager coefficients are symmetric, that is, $L_{k \ell} = L_{\ell k}$ for all $k, \ell$, even if the charges do not commute. This is therefore the non-Abelian generalization of Onsager's reciprocity relations. If the charges are not time-reversal invariant, as is the case, for instance, of spin chains, the matrix $\Lm$ is not necessarily symmetric, but we instead recover Onsager-Casimir reciprocity~\cite{Casimir1945} (see appendix \ref{app:O-C} for details).
 
Next we exploit the properties of the y-covariance to split the entropy production rate in Eq.~\eqref{Sigma_D} in classical and quantum contributions. For this purpose we  make use of the general relation $\covy(A,A) = \var(A) - I_y(\pi,A)$, where $\var(A) = \langle A^2 \rangle_\pi - \langle A\rangle_\pi^2$ is the conventional variance, and $I_y(\pi,A)$ is the so-called Wigner-Yanase-Dyson (WYD) skew information~\cite{Petz2002,Miller2019}, defined as 
\begin{equation}\label{WYD}
I_y(\pi,A) = - \frac{1}{2} \tr \Big( [\pi^y,A][\pi^{1-y},A]\Big).
\end{equation}
This quantity measures the amount of quantum coherence that $A$ has in the eigenbasis of $\pi$ {(or alternatively the coherence of $\pi$ in the $A$ eigenbasis)~\cite{Marvian2014}}. It is always non-negative and becomes zero if and only if $A$ commutes with $\pi$. Plugging this into the entropy production expression in Eq.~\eqref{Sigma_D}, we find
\begin{equation}\label{Sigma_sep}
{\Sigma} = \frac{1}{2} \var(D) - \frac{1}{2} \int\limits_0^1 dy \; I_y(\pi, D). 
\end{equation}
While for Abelian transport only the first term survives, the contribution of the WYD skew information, $I_y(\pi, D) \geq 0$, is to reduce the entropy production. This reduction is associated to the fluctuations in the operator $D$ {and is one of the main results of this paper. It shows how quantum coherence,  due to the non-commutativity of the thermodynamic charges, has a clear thermodynamic signature, effectively reducing the dissipation (entropy production) in the process. In particular, we see that the crucial ingredient for such a reduction is the quantum coherence between the observable $D$ associated to the currents and the global equilibrium state $\pi$.}
We introduce the relative reduction in entropy due to non-commutativity, which reads
\begin{equation} \label{eq:R}
 \mathcal{R} := \frac{1}{2} \int\limits_0^1 dy \; \frac{I_y(\pi, D)}{\Sigma} ~=~ \frac{1}{2} \frac{\langle D^2 \rangle_\pi}{\Sigma} - 1,
\end{equation}
where we used $\langle D \rangle_\pi = 0$ and hence $\mathrm{var}(D) = \langle D^2 \rangle_\pi$. The positivity of $I_y(\pi, D)$ and $\Sigma$ ensures $\mathcal{R} \geq 0$. This also leads to a universal upper bound to the entropy production, $\langle D^2 \rangle_\pi / 2 \geq \Sigma$, which is saturated in the classical case, when $\mathcal{R} = 0$.

The above results can be straightforwardly generalized to generic currents made by arbitrary combinations of the original charge currents in Eq.~\eqref{LR}, $J_k^\prime \equiv \sum_l a_{k l} J_l$, with $a_{k l}$ arbitrary real coefficients. In such case, the corresponding set of affinities $\lambda_k^\prime$ verify $\lambda_l = \sum_k a_{l k} \lambda_l^\prime$, leaving invariant the entropy production $\Sigma$ and the reduction $\mathcal{R}$ in Eq.~\eqref{eq:R}. The Onsager matrix for the new currents then reads $\Lm^\prime = \A \Lm \A^\dagger$, where $\A \equiv \{ a_{l k}\}$. 
As a consequence, we see that $\Lm^\prime$ continues to be  symmetric and positive semi-definite.

\begin{figure*}[t]
    \centering
    \includegraphics[width=1.0\textwidth]{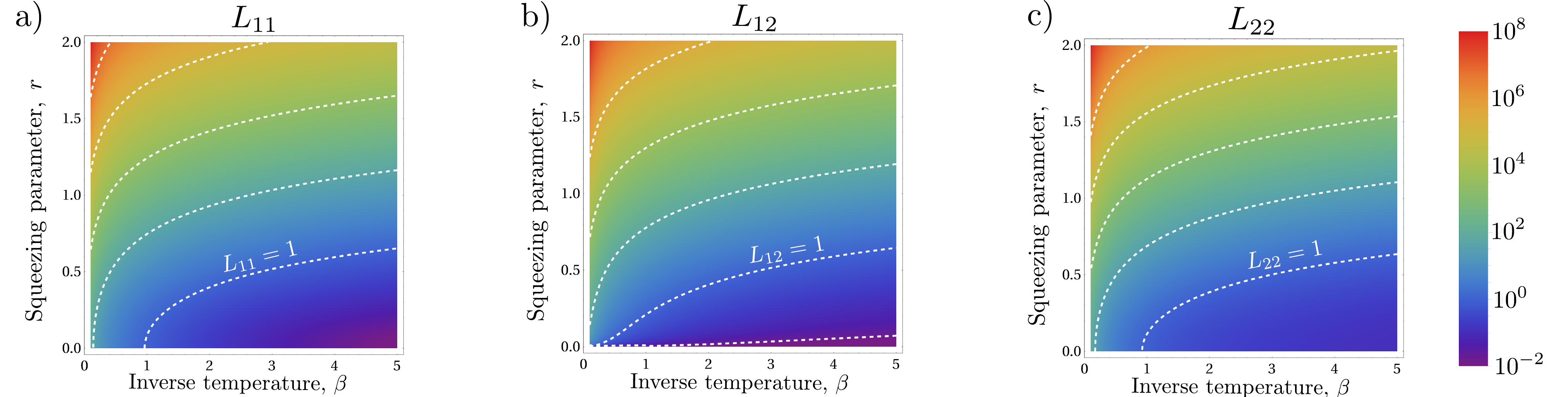}
    \caption{(a)-(c) Thermosqueezing Onsager Coefficients $L_{11}, L_{12}, L_{22}$ in logscale, computed from Eq.~\eqref{eq:1_energy}, in units of $(\hbar \omega)^2 \sin^2(g\tau)$, as a  function of the inverse temperature $\beta$ (in units of $\hbar \omega/k_B$) and the adimensional squeezing parameter $r$.}
    \label{fig:2}
\end{figure*}

\section{Thermosqueezing operations} \label{sec:thermosqueezing}
The general framework for Non-Abelian transport introduced here allow us to explore a broad range of situations in quantum thermodynamics, where coherence and thermal effects are reciprocally coupled. To illustrate this point, we now consider the joint transport of energy and squeezing in interacting bosonic systems prepared in squeezed thermal states~\cite{Manzano2016,Manzano2017b,Manzano2018b,Ronagel2014,Klaers2017a,Klaers2019}.

Let us focus first on the interaction between two resonant bosonic modes with frequency $\omega$, described by canonical variables $x_i, p_i$ satisfying $[x_i, p_j] = i \delta_{ij}$ (we set $\hbar = 1$). Here $i = 1,2$ refers to the two modes in question, which are prepared in a GGE of the form~\eqref{GGE} with charges $Q_1^{(i)}  \equiv H_i = \omega (p_i^2+x_i^2)/2$, associated to energy, and $Q_2^{(i)} \equiv A_i = \omega (p_i^2 - x_i^2)/2$, associated to squeezing~\cite{Manzano2016, Serafini2017}.
{These operators do not commute. In fact, together with $Q_3^{(i)} = \omega\{x_i,p_i\}/2$, they form a SU(1,1) non-Abelian group.
}
In analogy to the grand canonical ensemble, we identify the two affinities accompanying the charges as the inverse temperature $\lambda_1 \equiv \beta$ (associated to the Hamiltonian $H$) and $\lambda_2 \equiv - \beta \mu$ (associated to the squeezing asymmetry $A$), where $\mu$ is the equivalent of the chemical potential for squeezing currents (we do not include $Q_3^{(i)}$ among the charges of the GGE, merely for simplicity).
The corresponding GGE is thus nothing but a squeezed thermal state~\cite{Manzano2018b}, which can nowadays be prepared in many experimental platforms~\cite{Pirkkalainen2015,Wollman2015,Yurke1988}. The amount of squeezing is customarily quantified by the squeezing parameter $r$, related to the affinities by $\mu = \tanh(2 r) = \lambda_2/\lambda_1$. 

The interaction between the two modes must satisfy charge preservation, Eq.~\eqref{charge_preservation}. 
We show in appendix ~\ref{app:gaussian} that the only Gaussian unitary satisfying this property for two modes is a beam-splitter type unitary of the form 
\begin{equation}\label{beam_splitter}
U = e^{- g \tau (a_1^\dagger a_2 - a_2^\dagger a_1)},    
\end{equation}
where $a_i = (x_i + i p_i)/\sqrt{2}$ are the ladder operators of the bosonic modes, $g$ is the interaction strength and $\tau$ is the interaction time. To compute $\Lm$ we have developed a method based on the so-called Symmetric Logarithmic Derivative (SLD). 
We notice that the Onsager coefficients in Eq.~\eqref{onsager} can be rewritten as:
\begin{equation}\label{onsagerSLD}
 L_{k i} = \frac{1}{2} \langle \{Q_k^{(1)} - \xi(Q_k^{(1)}), \Lambda_i \} \rangle_{\pi^{(1)}},
\end{equation}
where $\xi(Q_k^{(1)}) \equiv \tr_2[\tilde{Q}_k^{(1)} ~\pi_{\bm \lambda}^{(2)}]$ and we introduced the SLD $\Lambda_i$, defined by:
\begin{equation}
\frac{1}{2} \left( \Lambda_i \pi_{\bm \lambda}^{(1)} + \pi_{\bm \lambda}^{(1)} \Lambda_i \right) \equiv \frac{\partial \pi^{(1)}_{\bm \lambda + d \bm \lambda}}{\partial (d \lambda_i)}\bigg|_{d \lambda_i = 0}.  
\end{equation}
The SLDs are hermitian and can be defined for either subsystem. In the case of Abelian transport (communting charges), they reduce to $\Lambda_i = \langle Q_i^{(1)} \rangle_{\pi^{(1)}} - Q_i^{(1)}$. For non-commuting charges the SLD can be calculated from the formula~\cite{Jiang2014}, valid for generic exponential quantum states:
\begin{equation} \label{SI_sldsol}
\Lambda_i 
= \langle Q_i^{(1)}\rangle_{\pi^{(1)}} - \sum\limits_{n=0}^\infty f_{2n} \mathcal{C}^{2n}\left( Q_i^{(1)} \right),
\end{equation}
where $\mathcal{C}^{m}(\mathcal{O}) \equiv [G, ... [G, \mathcal{O}]]$ is the $m$-th order nested commutator of an observable $\mathcal{O}$ with $G \equiv \sum_\ell \lambda_\ell Q_\ell$ [c.f. Eq.~\eqref{GGE}] and $\mathcal{C}^0(\mathcal{O}) = \mathcal{O}$. In the above expression $f_{m}= 4(4^{m/2 + 1} - 1) B_{m+2}/(m + 2)!$, where $B_{k}$ are the Bernouilli numbers.

By identifying $Q_1^{(1)} = H^{(1)}$ and $Q_2^{(1)} = A^{(1)}$ in Eq.~\eqref{onsagerSLD}, we have $\lambda_1= \beta$ and $\lambda_2= -\beta \mu$ and the SLD can be obtained by calculating the nested commutators (see appendix \ref{sec:SLD}). Using the SLD and the expressions for $Q_k^{(1)} - \xi(Q_k^{(1)})$, we arrive to closed formulas for the Onsager coefficients: 
 \begin{align} 
L_{11} &= \sin^2(g \tau) \frac{\omega^2}{1-\mu^2} [ \bar{n}^2 + \bar{n} + \mu^2 \left(\frac{\tanh(\alpha)}{\alpha}\right)(\bar{n}^2 + \bar{n}/2 + 1/2)], \nonumber \\
L_{12} &= \sin^2(g \tau) \frac{\omega^2 \mu}{1-\mu^2} [\bar{n}^2 + \bar{n} + \left(\frac{\tanh(\alpha)}{\alpha}\right)(\bar{n}^2 + \bar{n}/2 + 1/2)], \nonumber \\ \label{eq:1_energy}
L_{22} &= \sin^2(g \tau) \frac{\omega^2}{1-\mu^2} [\mu^2 (\bar{n}^2 + \bar{n}) +  \left(\frac{\tanh(\alpha)}{\alpha}\right)(\bar{n}^2 + \bar{n}/2 + 1/2)],
\end{align}
and $L_{21} = L_{12}$. Here $\bar{n} = (e^{\alpha} - 1)^{-1}$ and $\alpha = \beta \omega \sqrt{1 - \mu^2}$. For no squeezing ($\mu = 0$) the cross-coefficients vanish, $L_{12} = L_{2 1} = 0$. The above expressions can be simplified (but do not vanish) in both the high-temperature ($\beta \omega \ll 1$) or the high-squeezing ($\mu \rightarrow 1$) limits, where $\tanh(\alpha)/\alpha \rightarrow 1$.
These results are plotted in Fig.~\ref{fig:2}. Notably, we find that coefficients are monotonically decreasing functions of $\beta$, for fixed $\mu$, and monotonically increasing functions of $\mu$, for fixed $\beta$.
Note also that, in order to have non-zero transport, the unitary needs to be tuned to ensure $g \tau \neq n \pi$ for $n=0,1,2,...$ avoiding a complete SWAP.

\begin{figure*}[t]
\includegraphics[width= 1.0 \linewidth]{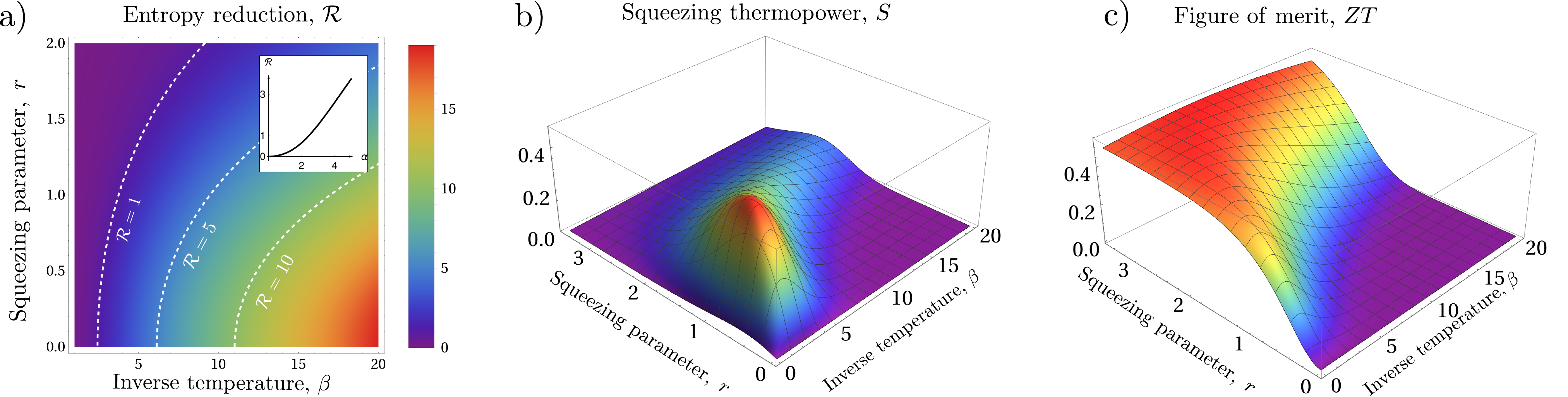}
\caption{(a) Entropy reduction $\mathcal{R}$, (b) squeezing thermopower coefficient $S$ and (c) figure of merit $ZT$ as a function of the inverse temperature $\beta$ and squeezing parameter $r$. The squeezing thermopower and $\beta$ are given in units of $\hbar \omega/k_B$ and $k_B/\hbar \omega$ respectively. Inset: Entropy reduction as a function of the adimensional parameter $\alpha = \beta \omega \sqrt{1 - \mu^2}$. In all plots we set $g \tau = \pi/2$.} \label{fig:3}
\end{figure*}

In turn, we can use these results to compute $\Sigma$ and, in particular, the entropy reduction $\mathcal{R}$ in Eq.~\eqref{eq:R}. The entropy production rate is given in terms of the Onsager coefficients in Eqs.~\eqref{eq:1_energy} as:
\begin{equation} \label{eq:sigmacoeffs}
 \Sigma = L_{11} \delta \lambda_1^2 + (L_{12} + L_{2 1}) \delta \lambda_1 \delta \lambda_2 + L_{22} \delta \lambda_2^2,  
\end{equation}
while the classical part of the entropy production is given from the variance of the operator $D$, as $\var(D) = \langle D^2 \rangle_{\pi^{(1)}}$ (see App.~\ref{sec:SLD}). Consequently, the relative entropy reduction $\mathcal{R}$ in Eq.~\eqref{eq:R} becomes:
\begin{align} \label{eq:Rsqueezing}
 \mathcal{R} = \frac{\alpha - \tanh(\alpha)}{2\alpha/(3\cosh(\alpha) - \sinh(\alpha) - 1) + \tanh(\alpha)},
\end{align}
which in the high-temperature and high-squeezing limits, $\alpha \ll 1$, vanishes $\mathcal{R} \rightarrow 0$.
Remarkably, we find that $\mathcal{R}$ does not depend on the affinity gradients $\delta \lambda_i$, but  it is only a function of the parameter $\alpha$ and hence only depends on the actual values of $\beta$ and $\mu$ (or $r$).
In fact, even more than that, it depends on $\beta$ and $\mu$ only via the parameter $\alpha = \beta \omega \sqrt{1 - \mu^2}$.
It can therefore be considered as a generic property of the setup. 
In Fig.~\ref{fig:3}(a) we show $\mathcal{R}$ as a function of  $\beta$ and  $\mu$ (and as a function of $\alpha$ in the inset). Interestingly, while greater entropy reductions are obtained at low temperatures (where quantum effects become dominant), we observe that a high squeezing in the reservoir actually  spoils the %quantum 
reduction effect.

\section{Thermosqueezing effects} \label{sec:effects}

In order to establish a connection with thermoelectric effects, we now focus on the Onsager coefficients for the transport of heat and squeezing. This can be done by noticing that heat current is a linear combination of energy and squeezing currents~\cite{Manzano2018b}, that is $J_Q^{(i)} \equiv J_1^{(i)} - \mu_i J_2^{(i)}$, while $J_A^{(i)} \equiv J_2^{(i)}$. The corresponding affinities are the changes in inverse temperatures and squeezing potentials, $\delta \lambda_1^\prime = \delta \beta$ and $\delta \lambda_2^\prime = - \beta \delta \mu - \mu \delta \beta$. In this case we obtain the following reciprocity relations (see appendix \ref{app:heat} for details):
\begin{align} \label{eq:reci}
 J_Q^{(1)} &= L_{QQ}~\delta \beta - L_{QA}~ \beta \delta \mu, \nonumber  \\
 J_A^{(1)} &= L_{AQ}~\delta \beta - L_{AA}~ \beta \delta \mu,
\end{align}
with Onsager coefficients ($\Lm^\prime \geq 0$) reading:
\begin{align} \label{eq:thermocoeffI}
L_{QQ} &= \omega^2 \sin^2(g \tau) \left(1 - \mu^2 \right)\bar{n}(\bar{n}+1), \nonumber \\  
L_{QA} &= \omega^2 \sin^2(g \tau) ~\mu~ \bar{n}(\bar{n}+1), \nonumber \\  
%\label{eq:thermocoeffF}
L_{AA} &= \omega^2 \sin^2(g \tau) \left(\frac{1}{1 - \mu^2} \right) \Big[ \mu^2\bar{n}(\bar{n}+1) \\ 
&~~+ \left(\frac{\tanh(\alpha)}{ \alpha}\right) (\bar{n}^2 + \bar{n}/2 + 1/2)\Big], \nonumber 
\end{align}
and $L_{AQ} = L_{Q A}$. In the absence of squeezing ($\mu \rightarrow 0$), the cross-coefficients vanish, $L_{QA}, L_{AQ} \rightarrow 0$. 
The behavior of these quantities with $\beta$ and $\mu$ is similar to those of Eq.~\eqref{eq:1_energy}: monotonically decreasing with $\beta$, and increasing with $\mu$.

The coefficients~\eqref{eq:thermocoeffI} can be  related to cross-effects linking transport in heat and squeezing, in analogy to thermoelectric coefficients~\cite{Callen1985}. We start by characterizing the coefficients $L_{QQ}$ and $L_{AA}$ by defining thermal and squeezing conductance. These relate heat and squeezing asymmetry currents to the corresponding gradients in temperature $\delta T = T_1 - T_2$ (at constant squeezing) and squeezing potentials $\delta \mu = \mu_1 - \mu_2$ (at constant temperature) as:
\begin{align} \label{eq:coeffs1}
\kappa &\equiv \Big(\frac{J_Q}{\delta T} \Big)_{\delta \mu = 0}= - \frac{L_{QQ}}{T^2}, \\  \label{eq:coeffs2}
G &\equiv \Big( \frac{J_A}{\delta \mu} \Big)_{\delta T = 0} = - \frac{L_{AA}}{T},
\end{align}
where $\kappa$ is the thermal conductance and $G$ is the analogous of the electric conductance for squeezing flows. We also used that $\delta \beta = - \delta T/T^2$. Using Eqs.~\eqref{eq:coeffs1} and \eqref{eq:coeffs2} the total heat dissipated in the setup~\cite{Benenti2017} reads
\begin{equation}
    \dot{Q} \equiv T \Sigma = \kappa \delta T^2/T + J_A^2/G.
\end{equation} 
The first term represents the heat dissipated by the thermal resistance, while the second is a Joule-like heating due to the squeezing current.

The \emph{Squeezing thermopower} (or \emph{Squeezing-Seebeck} coefficient) measures the squeezing potential difference $\delta \mu$ developed between reservoirs due to a difference in temperatures $\delta T$ when no net squeezing current is flowing:
\begin{equation} \label{eq:thermopower}
 S \equiv - \left( \frac{\delta \mu}{\delta T} \right)_{J_A = 0} = \frac{1}{T} \frac{L_{AQ}}{L_{AA}}.   
\end{equation}
Analogously, the \emph{Squeezing-Peltier} coefficient measures the heat current $J_Q$ produced by a squeezing current $J_A$ in isothermal conditions ($\delta T = 0$):
\begin{equation} \label{eq:peltier}
 \Pi \equiv \left( \frac{J_Q}{J_A} \right)_{\delta T = 0} = \frac{L_{QA}}{L_{AA}}.
\end{equation}
The symmetry of the coefficients leads to a relation between squeezing Seebeck and Peltier effects $\Pi = T S$.

A precise determination of the thermosqueezing coefficients, Eqs.~\eqref{eq:coeffs1}-\eqref{eq:peltier}, may be useful to a number of applications. For example, assuming the squeezing thermopower $S$ is known, Eq.~\eqref{eq:thermopower} allows the development of a squeezing-based thermometer. Closely following the principle on which thermocouples are based, temperature differences can be obtained from squeezing potential differences (or the other way around) through the relation $\delta T = - (1/S) \delta \mu$, valid when $J_A \rightarrow 0$.  This requires tuning the device to obtain zero squeezing current between the reservoirs. Possibilities to implement such "open-circuit" conditions may include introducing counter-terms in the interaction $U$ to effectively cancel the exchange of squeezing (e.g. applying local squeezing unitaries on each mode after the collision), or to interpose a small interface between the reservoirs blocking the squeezing current, while allowing the exchange of heat~\cite{noteJ}. In Fig.~\ref{fig:3}(b) we show the dependence of $S$ with respect to the temperature $\beta$ and squeezing parameter $r$. The small values of $S$ obtained suggest that tiny differences in squeezing potentials may be detected by measuring the larger temperature gradient.

The analogy with thermoelectricity also allow us to analyze this setup as a thermosqueezing engine. The squeezing current $J_A > 0$ can be used to power a heat flux $J_Q < 0$ against the temperature bias $\delta \beta > 0$, hence refrigerating the cold reservoir, when the difference in squeezing potentials is sufficiently high, $\delta \mu \geq \delta \mu_{\mathrm{fr}} \equiv  \kappa T \delta \beta/(GS)$ . The work (per collision) consumed by such fridge is given by $\dot{W} = J_A \delta \mu$, and can be made negative (work extraction) in the regime $0 \leq \delta \mu \leq \delta \mu_{\mathrm{stop}} \equiv \Pi (\delta \beta/\beta)$, where $J_Q > 0$. In the linear response regime, maximum power is obtained at $\delta \mu_\mathrm{stop}/2$, while maximum efficiency in both the refrigerator and engine regimes, as well as efficiency at maximum power, are determined by the adimensional figure of merit: 
\begin{equation}
ZT \equiv \frac{G S^2 T}{\kappa}, 
\end{equation}
defined in analogy to standard thermoelectrics~\cite{Esposito2009PRL, Thierschmann2015, Benenti2017}.
In Fig.~\ref{fig:3}(c) we show $ZT$ as a function of the squeezing parameter for different values of the temperature. It quickly saturates when increasing the squeezing parameter at any temperature, leading to values $ZT \simeq 0.5$, corresponding to $\eta(\dot{W}_{\mathrm{max}}) \simeq \eta_C/10$, where $\eta_C \equiv 1-\beta_1/\beta_2$ is the Carnot efficiency. 
This result can be intuitively understood from the fact that, since $[H,A] \neq 0$, heat and asymmetry currents can never be proportional to each other, hence avoiding the so called \emph{tight coupling} condition~\cite{Esposito2009PRL, Thierschmann2015, Benenti2017}.

\section{Conclusions}
We formulated a general framework based on a collisional model, and derived Onsager reciprocity relations for non-Abelian transport in linear-response. Our results lead to the identification of a entropy reduction effect induced by {quantum coherence and the non-commutativity of thermodynamic charges}. We illustrated our findings by developing reciprocal relations between energy and squeezing in bosonic setups. The applications of such reciprocal theory may be of use to a variety of quantum information technologies and enable new ways to convert heat into work in the quantum realm.

Our results are within reach of current experimental devices employed in thermodynamic scenarios, like the trapped ion platform used in Ref.~\cite{Maslennikov2019}, where different motional modes can be weakly coupled through Raman beams and prepared in squeezed thermal states reaching values up to $r \sim 1.5$. Another option would be building upon the squeezing-based nanobeam engine in Ref.~\cite{Klaers2017a}, able to reach squeezings around $r\sim 0.5$ at room temperatures. The quantum limit can be instead achieved in other setups for micromechanical oscillators by using reservoir-engineering techniques like two-tone driving~\cite{Pirkkalainen2015,Wollman2015}.
A third possibility would be to consider setups combining optical cavities and ultra-cold atoms~\cite{Brooks2012,Leonard2017a}. In \cite{Brooks2012} ponderomotive squeezing has been achieved in optical modes using an atomic cloud, while in \cite{Leonard2017a} a Bose gas interacts simultaneously with two optical cavities. Coupling the device to an extra cavity in the former case, and by appropriately pumping the cavities with squeezed radiation in the later, one could effectively implement a thermosqueezing device and thus also explore some of the consequences presented in this framework.

Extensions to situations with multiple terminals and including external modulation may be important in view of applications, in close analogy with thermoelectricity~\cite{Mazza2014, Bradner2013, Sothmann2014}. Of particular interest would be to consider squeezing in the context of hybrid thermal machines operating with multiple conserved quantities~\cite{Manzano2020}, as well as the study of fluctuations in the currents for non-Abelian setups, where standard exchange fluctuation theorems break down~\cite{Andrieux2009,Campisi2010}. Finally, our theory should also lead to fruitful applications in other non-Abelian quantum systems~\cite{YungerHalpern2020, Ikeda2020}.

\begin{acknowledgments}
We thank J.P. Santos for fruitful discussions. 
The authors acknowledge the support from the Abdus Salam International Centre of Theoretical Physics, where part of this work was developed, for both the hospitality and the financial support. G.M. acknowledges funding from Spanish MICINN through the Juan de la Cierva program (IJC2019-039592-I) and from the European Union's Horizon 2020 research and innovation program under the Marie Sk\l{}odowska-Curie grant agreement No. 801110 and the Austrian Federal Ministry of Education, Science and Research (BMBWF). 
J.M.R.P. acknowledges financial support from the Spanish Government (Grant Contract, FIS-2017-83706-R) and from the Foundational Questions Institute Fund, a donor-advised fund of Silicon Valley Community Foundation (Grant number FQXi-IAF19-01).
G.T.L. acknowledges the hospitality of Apt. 44, where part of this work was developed, and the financial support of the S\~ao Paulo Funding Agency FAPESP (Grants No. 2017/50304-7, 2017/07973-5 and 2018/12813-0) and the Brazilian funding agency CNPq (Grant No. INCT-IQ 246569/2014-0).
\end{acknowledgments}

\appendix

\section{Informational form of the entropy production rate} \label{sm:1}

Equation~\eqref{Pi_gen} in Sec.~\ref{sec:framework} expresses the entropy production rate per collision in terms of the charge fluxes and thermodynamic forces. Here we show that such expression is equivalent to the fully informational form for entropy production derived in Refs.~\cite{Manzano2017a,Santos2019,Strasberg2016,Esposito2010a} and show its positivity. Following the informational approach, the total entropy production due to a single collision can be written as:
\begin{eqnarray}\label{SM_sigma}
\Sigma &= \Delta S_1 + \Delta S_2 + S(\rho_1' || \pi_{\bm{\lambda}^{(1)}}^{(1)})+ S(\rho_2' || \pi_{\bm{\lambda}^{(2)}}^{(2)}) \nonumber \\ &= \mathcal{I}(\rho_{12}') 
+ S(\rho_1' || \pi_{\bm{\lambda}^{(1)}}^{(1)})+ S(\rho_2' || \pi_{\bm{\lambda}^{(2)}}^{(2)}) \geq 0,
\end{eqnarray}
where $\Delta S_i = S(\rho_i') - S(\pi_{\bm{\lambda}^{(i)}}^{(i)})$ represents the change in (local) von Neumann entropy of subsystem $i = 1,2$ due to the interaction, and $S(\rho|| \sigma) = \tr(\rho \ln\rho - \rho \ln \sigma)$ is the quantum relative entropy, accounting for the entropic cost necessary for reseting the subsystems back to equilibrium, i.e. for maintaining the reservoirs in equilibrium~\cite{Manzano2017a}. In the second equality  we introduced the mutual information $\mathcal{I}(\rho_{12}') = S(\rho_1') + S(\rho_2') - S(\rho_{12}')$ developed between systems 1 and 2 due to the map in Eq.~(2). Since the initial state is a product state and the interaction is unitary, it follows that $\mathcal{I}(\rho_{12}') = \Delta S_1 + \Delta S_2$. By expressing the entropy production in the form~\eqref{SM_sigma} the 2nd law is automatically satisfied,  $\Sigma \geq 0$, since each  term in~\eqref{SM_sigma} is individually non-negative.

Using the explicit expressions for the entropy changes $\Delta S_i = S(\rho_i') - S(\pi_{\bm{\lambda}^{(i)}}^{(i)})$ and the relative entropy contributions $S(\rho_i' || \pi_{\bm{\lambda}^{(i)}}^{(i)}) = \tr(\rho_i' \ln\rho_i' - \rho \ln \pi_{\bm{\lambda}^{(i)}}^{(i)})$, we arrive to the form
\begin{equation} \label{SM_sigma2}
\Sigma = \tr[(\pi_{\bm{\lambda}^{(1)}}^{(1)} - \rho_1') \ln _{\bm{\lambda}^{(1)}}^{(1)}] + 
\tr[(\pi_{\bm{\lambda}^{(2)}}^{(2)} - \rho_2') \ln _{\bm{\lambda}^{(2)}}^{(2)}],
\end{equation}
which, upon substituting for the GGE in Eq.~\eqref{GGE}, leads exactly to Eq.\eqref{Pi_gen}. Finally, we stress that, since every collision starts with the same product of initial GGE states $\pi_{\bm{\lambda}^{(1)}}^{(1)} \otimes \pi_{\bm{\lambda}^{(2)}}^{(2)}$ and is subjected to the same unitary interaction $U$, the expression in Eq.~\eqref{SM_sigma2} represents the entropy production rate (per collision).

\section{Proof of Theorem I} \label{sm:2}

In this appendix we provide a detailed proof of Theorem I. Recall we are assuming that $\lambda_k^{(2)} = \lambda_k$ but $\lambda_k^{(1)} = \lambda_k + \delta \lambda_k$, where $\delta \lambda_k$ are small real numbers. We use the following Feynman integral representations, valid for an arbitrary operator $G(\phi)$, depending on some parameter $\phi$:
\begin{equation}
\partial_\phi e^{-G} 
= -\int\limits_0^1 dy \; e^{- G y} 
(\partial_\phi G)
e^{-G(1-y)} .
\end{equation}
Applying this to $\pi^{(1)}$, as given in Eq.~\eqref{GGE}, we find
\[
\frac{\partial \pi_{\bm{\lambda} + \delta \bm{\lambda}}^{(1)}}{\partial (\delta \lambda_\ell)} \bigg|_{\delta\lambda_\ell = 0} 
 = \langle Q_\ell^{(1)} \rangle_{\pi^{(1)}}~\pi^{(1)}
- \int\limits_0^1 dy \; (\pi_{\bm{\lambda}}^{(1)})^y Q_\ell^{(1)} (\pi_{\bm{\lambda}}^{(1)})^{1-y},
 \]
where we also used that
\[
-\frac{\partial \ln Z_1}{\partial (\delta \lambda_\ell)}\bigg|_{\delta\lambda_\ell = 0} = \langle Q_\ell^{(1)}\rangle_{\pi}.
\]
The above expressions can be rewritten by taking the tensor product with $\pi_{\bm{\lambda}}^{(2)}$. We then get 
\begin{align} \label{SM_expansion}
\frac{\partial \pi_{\bm{\lambda} + \delta \bm{\lambda}}^{(1)}}{\partial (\delta \lambda_\ell)} \bigg|_{\delta\lambda_\ell = 0} \otimes \pi_2
&=  \langle Q_\ell^{(1)} \rangle_{\pi}~\pi
- \int\limits_0^1 dy \; \pi^y Q_\ell^{(1)} \pi^{1-y} \nonumber \\
&=  \langle Q_\ell^{(1)} \rangle_{\pi}~\pi
- \int\limits_0^1 dy \; \pi^{1-y} Q_\ell^{(1)} \pi^{y},
\end{align}
where $\pi = \pi_{\bm{\lambda}}^{(1)} \otimes \pi_{\bm{\lambda}}^{(2)}$ and, in the second equality, we simply changed variables from $y$ to $1-y$.
We write these two forms here, side-by-side, as they will be useful in understanding the conditions for $L_{k\ell}$ to be symmetric. 

Inserting Eq.~\eqref{SM_expansion} in Eq.~\eqref{SM_L_original} leads to the two forms for the Onsager matrix, 
 \begin{align} \label{SM_Lkl_middle_1}
L_{k\ell} 
&=
- \int\limits_0^1 dy \; \tr [\big( \tilde{Q}_k^{(1)} - Q_k^{(1)} \big) \pi^y Q_\ell^{(1)} \pi^{1-y}]  \\ \label{SM_Lkl_middle_2}
&= 
- \int\limits_0^1 dy \; \tr [\big( \tilde{Q}_k^{(1)} - Q_k^{(1)} \big) \pi^{1-y} Q_\ell^{(1)} \pi^{y} ],
\end{align}
where we used the fact that terms of the form $\langle Q_\ell^{(1)} \rangle_{\pi} \langle \tilde{Q}_k^{(1)} - Q_k^{(1)} \rangle_{\pi   }$ vanish since $U\pi U^\dagger = \pi$.

%%%%%%%%%%%%%%%%%%%%%%%

Eqs.~\eqref{SM_Lkl_middle_1}-\eqref{SM_Lkl_middle_2} provide two equivalent ways of writing the Onsager coefficients. We now prove that when the charges, and the dynamics, are time-reversal invariant, the matrix $L$ is symmetric; i.e., $L_{\ell k} = L_{k\ell}$. To accomplish this, we essentially exchange $k\leftrightarrow \ell$ in the first equality in Eq.~\eqref{SM_Lkl_middle_1} and compare with the second one. Indeed, exchanging $k\leftrightarrow \ell$ and using the cyclic property of the trace, we obtain
\begin{equation}\label{SM_tmp4531321}
L_{\ell k} = 
- \int\limits_0^1 dx \; \tr [ Q_k^{(1)} \pi^{1-x} \tilde{Q}_\ell^{(1)} \pi^x ]
+ \int\limits_0^1 dx \; \tr [ Q_k^{(1)} \pi^{1-x} Q_\ell^{(1)} \pi^x ].
\end{equation}
We notice that the second term is already the same as the last term of the second equality in Eq.~\eqref{SM_Lkl_middle_2}. 
We thus need to focus only on the first term. Since $\pi$ is a fixed point of $U$ we can write the first term above as 
\begin{equation}\label{SM_inv}
\tr [ Q_k^{(1)} \pi^{1-x} \tilde{Q}_\ell^{(1)} \pi^x ] = 
\tr [ U Q_k^{(1)} U^\dagger \pi^{1-x} Q_\ell^{(1)} \pi^x ],
\end{equation}
where we stress that $U Q_k^{(1)} U^\dagger \neq \tilde{Q}_k^{(1)}$. We now introduce the time-reversal operator in quantum mechanics $\Theta$, responsible of changing the sign of odd variables under time inversion, such as momenta and magnetic field~\cite{Andrieux2008,Haake2010}. Using it we can define the time-reversal of the unitary evolution $\bar{U} = \Theta U^\dagger \Theta^\dagger$ and time-reversal of the charges as $\bar{Q}_k^{(1)} = \Theta Q_k^{(1)} \Theta^\dagger$. 
Introducing these operators in the above equation~\eqref{SM_inv}, it then follows that 
\begin{equation}\label{SM_timer}
\tr [ U Q_k^{(1)} U^\dagger \pi^{1-x} Q_\ell^{(1)} \pi^x ]
=
\tr [ \bar{U}^\dagger \bar{Q}_k^{(1)} \bar{U} \Theta \pi^{1-x} \Theta^\dagger \bar{Q}_\ell^{(1)} \Theta \pi^x \Theta^\dagger ].
\end{equation}
If the dynamics is time-reversal invariant, then $\bar{U} = U$ is verified. Moreover, if the charges are time-reversal invariant then $\bar{Q}_k^{(1)} = Q_k^{(1)}$ and $\Theta \pi\Theta = \pi$. Using this in Eq.~\eqref{SM_timer}, we finally arrive at 
\begin{equation}\label{SM_final}
\tr [ Q_k^{(1)} \pi^{1-x} \tilde{Q}_\ell^{(1)} \pi^x ] = 
\tr [ \tilde{Q}_k^{(1)} \pi^{1-x} Q_\ell^{(1)} \pi^x ].
\end{equation}
Plugging Eq.~\eqref{SM_final} in Eq.~\eqref{SM_tmp4531321} and comparing with~\eqref{SM_Lkl_middle_2} then clearly shows that $L_{\ell k} = L_{k\ell}$. 

%%%%%%%%%%%%%%%%%%%%%%%

We finally express the results in terms of the $y$-covariance, defined in Eq.~\eqref{covy} of Sec.~\ref{sec:theorem}. Comparing Eq.~\eqref{covy} with~\eqref{SM_Lkl_middle_1} immediately shows us that we can write
\begin{equation}
L_{k\ell} = \int\limits_0^1 dy \; \covy(Q_k^{(1)} - \tilde{Q}_k^{(1)}, Q_\ell^{(1)}).
\end{equation} 
Now using that $L$ is symmetric, we can also write the above equation as an average between $L_{k\ell}$ and $L_{\ell k}$, leading to
\[
	L_{k\ell} = \frac{1}{2}(L_{k \ell} + L_{\ell k}) = \frac{1}{2} \int\limits_0^1 dy \; \covy\Big(\tilde{Q}_k^{(1)} - Q_k^{(1)}, \tilde{Q}_\ell^{(1)}  - Q_\ell^{(1)}\Big),
\]
which is Eq.~\eqref{onsager} in Theorem I. $\square$

\section{Onsager-Casimir reciprocity}\label{app:O-C}
As stressed in Sec.~\ref{sec:theorem}, Onsager reciprocity follows from stationarity of the dynamics, $U\pi U^\dagger = \pi$, \emph{and} invariance under time reversal, $\bar{U}=U$ and $\bar{Q}=Q$. However, the second assumption can be relaxed to obtain a weaker form of reciprocity, named Onsager-Casimir reciprocity~\cite{Casimir1945}. 

In this case time-inversion of the dynamics leads to a different evolution, $\bar{U}=U_\ast$, and potentially different charges $\bar{Q}_k^{(1)} = Q_k^{\ast (1)}$, depending on the parity of the operators $Q_k^{(1)}$. In the evolution $U_\ast$ and charges $Q_k^{\ast (1)}$ external magnetic fields and/or other odd variables change their signs. Using these relations in  Eq.~\eqref{SM_timer} above, it follows that:
\begin{equation}\label{SM_casimir}
\tr [ Q_k^{(1)} \pi^{1-x} \tilde{Q}_\ell^{(1)} \pi^x ] = \tr [ \tilde{Q}_k^{\ast (1)} \pi_\ast^{1-x} Q_\ell^{\ast(1)} \pi_\ast^x ],
\end{equation}
where $\tilde{Q}_k^{\ast (1)} = U_\ast^{\dagger} Q_\ell^{\ast(1)} U_\ast$ and $\pi_\ast = e^{-\sum_k \lambda_k Q_k^\ast}/Z_\ast = \pi_{\bm{\lambda}}^{\ast (1)} \otimes \pi_{\bm{\lambda}}^{\ast (2)}$ is the fixed point of the evolution, $U_\ast \pi_\ast U_\ast^\dagger = \pi_\ast$. Introducing the time-reversal operator in the second term of Eq.~\eqref{SM_tmp4531321} we then arrive to:
\begin{equation}
 L_{\ell k} = L_{k \ell}^\ast,
\end{equation}
where the transport coffieints $L_{k \ell}^\ast$ are defined similarly  to $L_{k \ell}$, but with respect to the modified evolution $U_\ast$ and charges $Q_k^{\ast (1)}$ in the GGE $\pi_{\bm{\lambda} + \delta\bm{\lambda}}^{\ast (1)}$; that is:
\begin{equation}\label{SM_L_casimir}
L_{k\ell}^\ast = \tr[ (\tilde{Q}_k^{\ast (1)} - Q_k^{\ast (1)}) \left(\frac{\partial \pi_{\bm{\lambda} + \delta \bm{\lambda}}^{\ast (1)}}{\partial (\delta \lambda_\ell)} \bigg|_{\delta\lambda_\ell = 0} \otimes \pi_{\bm{\lambda}}^{\ast (2)}\right)].
\end{equation}

\section{General Gaussian thermosqueezing operations} \label{app:gaussian}
 In this appendix we discuss the general structure of thermosqueezing operations which preserve Gaussianity. We assume systems 1 and 2 are composed of an arbitrary number $N_1$ and $N_2$ of bosonic modes. Let $\mathcal{N}_\alpha$ denote the set of modes belonging to system $\alpha = 1,2$. We then define the net charges $Q_{1,2,3}$ associated to each system as 
\begin{IEEEeqnarray}{rCl}
\nonumber
Q_1^{(\alpha)} = \frac{\hbar \omega}{2}\sum\limits_{i\in\mathcal{N}_\alpha} (p_i^2 +  x_i^2) &~~;~~& Q_2^{(\alpha)} = \frac{\hbar \omega}{2} \sum\limits_{i\in\mathcal{N}_\alpha} (p_i^2 - x_i^2) \\ %~~~;~~~ 
Q_3^{(\alpha)} = &\frac{\hbar \omega}{2}& \sum\limits_{i\in\mathcal{N}_\alpha} \{x_i ,p_i \}.
\end{IEEEeqnarray}
We are interested here in Gaussian operations satisfying the charge preservation condition~\eqref{charge_preservation}. 
Let $\bm{R} = (x_1,x_2,\ldots, p_1,p_2,\ldots)$ denote a vector of size $2N$ (where $N = N_1 + N_2$) containing all position operators of systems 1 and 2, followed by all momentum operators of 1 and 2. 
A generic Gaussian unitary can then be translated into the symplectic transformation 
\begin{equation}\label{SI_unitary_R_action}
U^\dagger R_i U = \sum\limits_j V_{ij} R_j, 
\end{equation}
where $V$ is a $2N$-dimensional symplectic matrix, satisfying 
\begin{equation}\label{SI_general_thermo_symplectic}
V^\text{T} \Omega V  = \Omega,
\end{equation}
with $\Omega = \begin{pmatrix} 0 & \id_N \\ -\id_N & 0 \end{pmatrix}$ being the symplectic form and $\id_N$  the identity of dimension $N$. 

Our goal now is to obtain the additional restrictions (besides being symplectic) that are imposed on $V$ if it is to satisfy the charge preservation condition~\eqref{charge_preservation}. We begin by rewriting Eq.~\eqref{charge_preservation} as
\begin{equation}\label{SI_general_thermo_charge_preservation}
U^\dagger \big( Q_i^{(1)} + Q_i^{(2)} \big) U = Q_i^{(1)} + Q_i^{(2)}.
\end{equation}
Next we note that, since $Q_i^{(\alpha)}$ are quadratic in the vector $\bm{R}$, we can rewrite them as
\[
Q_i^{(1)} + Q_i^{(2)} = \frac{\hbar\omega}{2} \sum\limits_{m,n} (K_i)_{mn} R_m R_n, 
\]
where $K_i$ for $i=1,2,3$ are $2N$-dimensional matrices reading 
\begin{equation}
K_1 = \id_{2N}, ~~~~
K_2 = \begin{pmatrix} \id_N & 0 \\[0.2cm] 0 & -\id_N \end{pmatrix}, ~~~
K_3 = \begin{pmatrix} 0 & \id_N \\[0.2cm] \id_N & 0 \end{pmatrix}.
\end{equation}
Using Eq.~\eqref{SI_unitary_R_action}, we may now recast 
Eq.~\eqref{SI_general_thermo_charge_preservation} in the form 
\begin{equation}\label{SI_general_thermo_charge_preservation_symplectic}
V^\text{T} K_i V = K_i. 
\end{equation}
This can be viewed as the symplectic counterpart of the Hilbert space identity~\eqref{SI_general_thermo_charge_preservation}.
Together with~\eqref{SI_general_thermo_symplectic}, it yields a total of 4 restrictions on the form of $V$. 

Energy preservation is related to $K_1$. 
As a consequence,  in addition to being symplectic, Eq.~\eqref{SI_general_thermo_charge_preservation_symplectic} determines that $V$ must also be orthogonal; that is,  $V^\text{T} V = \id_{2N}$. 
As shown e.g. in Ref.~\cite{Dutta1995}, the intersection of the symplectic and orthogonal groups are matrices of the form 
\begin{equation}
O(2N) \cap \text{Sp}(2N) = \Bigg\{ \begin{pmatrix} X & Y \\ -Y & X \end{pmatrix} \;\big| \; X - i Y = U(N) \Bigg\}, 
\end{equation}
where $U(N)$ is the group of unitary matrices. 
All symplectic forms preserving $Q_1$ must therefore be of this form. 

On top of this, we now impose the conditions in Eq.~\eqref{SI_general_thermo_charge_preservation_symplectic} for preservation of squeezing,  $Q_2$ and/or $Q_3$. As one can verify, these will only be satisfied if $Y = 0$, independently of whether we impose the conservation only of $Q_2$, $Q_3$, or both of them. 
We thus finally conclude that the most general thermosqueezing operation has the form 
\begin{equation}\label{SI_thermosqueezing_general_final_form}
V = \begin{pmatrix} X & 0 \\[0.2cm] 0 & X \end{pmatrix},
\qquad X^\text{T} X = 1.
\end{equation}
The matrix $X$ must be real and also a member of $U(N)$, which is tantamount to $X\in O(N)$, the group of orthogonal matrices of size $N$. 
We remark that Eq.~\eqref{SI_thermosqueezing_general_final_form} is written with respect to the ordering $\bm{R} = (x_1,x_2,\ldots, p_1,p_2,\ldots)$. 
If one wishes to use the more standard ordering  $\tilde{\bm{R}} = (x_1,p_1, x_2,p_2,\ldots)$, then the entries of $V$ must be rearranged appropriately. 
 
The particular case employed in our example, Eq.~\eqref{beam_splitter}, corresponds to $N_1 = N_2 = 1$; that is, each system is composed of only a single mode. In this case the matrix $X$ will be of size $N = N_1 + N_2 = 2$ and hence can be parametrized as 
\begin{equation}
X = \begin{pmatrix}
\cos\phi & -\sin\phi \\[0.2cm]
\sin\phi & \cos\phi
\end{pmatrix},
\end{equation}
which is quantified by a single parameter $\phi$. This corresponds precisely to the unitary 
\begin{equation}\label{SI_Unitary_form}
U = e^{-i g \tau (x_1 p_2 - x_2 p_1)} = e^{- g \tau (a_1^\dagger a_2 - a_2^\dagger a_1)},
\end{equation}
with the identification  $\phi = g \tau$, $g$ being the interaction strength and $\tau$ is the interaction time. Interestingly, this analysis also shows that the unitary~\eqref{SI_Unitary_form} is the \emph{only} 2-mode Gaussian unitary preserving both energy and squeezing.

We also call attention to the phase appearing in Eq.~\eqref{SI_Unitary_form}. 
An interaction of the form $e^{-i g (a_1^\dagger a_2 + a_2^\dagger a_1)}$, for instance, conserves energy, but does not conserve squeezing. 
What the results above show is that $i(a_1^\dagger a_2 - a_2^\dagger a_1)$ is the only type of interaction conserving both energy \emph{and} squeezing.

\section{Onsager coefficients for energy and squeezing} \label{sec:SLD}

In this appendix we give details on the calculation of the Onsager matrix $\Lm$ using the SLD, as well as the contributions to the entropy production. As mentioned in Sec.~\ref{sec:thermosqueezing}, identifying $Q_1^{(1)} = H^{(1)}$ and $Q_2^{(1)} = A^{(1)}$ in Eq.~\eqref{onsagerSLD}, yields to $\lambda_1= \beta$ and $\lambda_2= -\beta \mu$. Using Eq.~\eqref{SI_sldsol} and calculating the nested commutators we arrive at:
\begin{align}
% \label{eq:lambdas1}
 \Lambda_1 &= \langle H \rangle_{\pi^{(1)}} - \left(H + \left[ \frac{\tanh(\alpha)}{\alpha} - 1 \right] \frac{\mu}{1-\mu^2}(A - \mu H)\right) \nonumber \\ \label{eq:lambdas2}
 \Lambda_2 &= \langle A \rangle_{\pi^{(1)}} - \left(A + \left[ \frac{\tanh(\alpha)}{\alpha} - 1 \right] \frac{1}{1-\mu^2}(A - \mu H)\right),
\end{align}
where we used $\sum_{n=1}^\infty f_{2n} \alpha^{2n} = \tanh(\alpha)/\alpha$ and we recall that $\alpha = \beta \hbar \omega \sqrt{1 - \mu^2}$. In the high-temperature ($\beta \omega \ll 1$) or high-squeezing ($\mu \rightarrow 1$) limits, we have $\alpha \ll 1$ and hence $\tanh(\alpha)/\alpha \rightarrow 1$. In such case we recover the expressions of the SLD in the Abelian case.

For the quantities $Q_k^{(1)} - \tilde{Q}_k^{(1)}$ we obtain:
\begin{align}\label{eq:changes}
 Q_1^{(1)} - \tilde{Q}_1^{(1)} &=  \sin^2(g\tau) (H^{(1)} - H^{(2)}) \nonumber \\ &~~- \cos(g \tau) \sin(g \tau) \omega (a_1^\dagger a_2 + a_1 a_2^\dagger), \nonumber \\
 Q_2^{(1)} - \tilde{Q}_2^{(1)} &=  \sin^2(g\tau) (A^{(1)} - A^{(2)}) \nonumber \\ &~~- \cos(g \tau) \sin(g \tau) \omega (a_1 a_2 + a_1^\dagger a_2^\dagger).
\end{align}
Here $a_i$ for $i=1,2$ are the ladder operators of the two bosonic modes. Taking the trace over mode 2 in the equilibrium state $\pi_{\bm \lambda}^{(2)}$ leads to:
\begin{align} 
%\label{eq:expre1}
 Q_1^{(1)} - \xi({Q}_1^{(1)}) &=  \sin^2(g\tau) (H^{(1)} - \langle H^{(1)}\rangle_{\pi^{(1)}}), \nonumber \\ \label{eq:expre2}
 Q_2^{(1)} - \xi({Q}_2^{(1)}) &=  \sin^2(g\tau) (A^{(1)} - \langle A^{(1)} \rangle_{\pi^{(1)}}),
\end{align}
where we took advantage of the fact that in equilibrium $\langle Q_k^{(1)}\rangle_{\pi^{(1)}} = \langle Q_k^{(2)}\rangle_{\pi^{(2)}}$.

Introducing the SLDs in Eqs.~\eqref{eq:lambdas2} and the expressions \eqref{eq:expre2} into Eq.~\eqref{onsagerSLD} we arrive to the Onsager coefficients in Eq.~\eqref{eq:1_energy}. In order to arrive at that final form, we also used the relations:
\begin{align} \label{eq:relations}
 \langle H^2 \rangle_\pi &= \cosh^2(2r) \langle H^2 \rangle_{\rm th} + \sinh^2(2r) \langle A^2 \rangle_{\rm th}, \nonumber \\
 \langle A^2 \rangle_\pi &= \cosh^2(2r) \langle A^2 \rangle_{\rm th} + \sinh^2(2r) \langle H^2 \rangle_{\rm th}, \nonumber \\
 \langle H A \rangle_\pi &= \langle A H \rangle_\pi = \sinh(2r) \cosh(2r) (\langle H^2 \rangle_{\rm th} + \langle A^2 \rangle_{\rm th}), \nonumber \\
 \langle H \rangle_\pi^2 &= \cosh^2(2r) \langle H^2 \rangle_{\rm th}^2, \nonumber \\
 \langle A \rangle_\pi^2 &= \sinh^2(2r) \langle H^2 \rangle_{\rm th}^2, \nonumber \\
 \langle H \rangle_\pi \langle A \rangle_\pi &= \sinh(2r) \cosh(2r) (\langle H \rangle_{\rm th}^2 + \langle A \rangle_{\rm th}^2),
\end{align}
where $\langle Q_i \rangle_{\rm th} = \tr[Q_i e^{-\alpha H/\omega}/Z_\alpha]$ are thermal averages in a Gibbs state at temperature $\alpha/\omega$ and we omitted the superscripts $(i)$ in $H$, $A$ and $\pi$ since the expressions are the same for modes 1 or 2. The explicit expresions for the thermal averages introduced above are:
\begin{align} \label{eq:relations2}
 \langle H^2 \rangle_{\rm th} &= (\hbar \omega)^2 (2 \bar{n}^2 + 2 \bar{n} + 1/4), \nonumber \\
 \langle A^2 \rangle_{\rm th} &= (\hbar \omega)^2 (\bar{n}^2 + \bar{n}/2 + 1/2), \nonumber \\
 \langle H \rangle_{\rm th}^2 &= (\hbar \omega)^2 (\bar{n} + 1/2)^2.
\end{align}

The classical part of the entropy production can be calculated from the operator $D = \sum_k \delta_k (\tilde{Q}_k^{(1)} - {Q}_k^{(1)})$ as:
\begin{align}
 &\var(D) = \langle D^2 \rangle_{\pi^{(1)}}  \\
 &= 2 \sin^2(g \tau) \Big( \delta \lambda_1^2 [\cosh^2(2r) (\bar{n}^2 + \bar{n}) + \sinh^2(2r) (\bar{n}^2 + \bar{n}/2 + 1/2)] \nonumber \\
 &+ 2 \delta \lambda_1 \delta \lambda_2 \cosh^2(2r) \sinh^2(2r)[(\bar{n}^2 + \bar{n}) + (\bar{n}^2 +\bar{n}/2 + 1/2)] \nonumber \\
 &+ \delta \lambda_2^2[\sinh^2(2r)(\bar{n}^2 + \bar{n}) + \cosh^2(2r)(\bar{n}^2 + \bar{n}/2 + 1/2)] \Big) \nonumber
\end{align}
where we used the expressions obtained above in Eqs.~\eqref{eq:changes} and the relations~\eqref{eq:relations} and \eqref{eq:relations2}. The relative entropy reduction $\mathcal{R}$ in Eq.~\eqref{eq:Rsqueezing} is then obtained by plugging the above expression and the entropy production rate $\Sigma$ [Eq.~\eqref{eq:sigmacoeffs}] into the definition of $\mathcal{R}$ in Eq.~\eqref{eq:R}.
 
\section{Onsager coefficients for heat and squeezing} \label{app:heat}
 
In order to calculate the Onsager coefficients for the currents of heat and squeezing asymmetry, we proceed as in the previous subsection using expression \eqref{onsagerSLD}. However we now identify as charges $Q_1^{\prime (1)} = H^{(1)} - \mu A^{(1)}$ and $Q_2^{\prime (1)} = Q_2^{(1)} = A^{(1)}$, and affinity gradients $\delta \lambda_1'= \delta \beta$ and $\lambda_2' = - \beta \delta \mu$, leading to the reciprocity relations in Eq.~\eqref{eq:reci}. In this case we obtain for the SLDs:
\begin{align} \label{eq:lambdas1p}
 \Lambda_1^\prime &= \langle H \rangle_{\pi^{(1)}} - \mu \langle A \rangle_{\pi^{(1)}} -  H +  \mu A   \\ \label{eq:lambdas2p}
 \Lambda_2^\prime &= \langle A \rangle_{\pi^{(1)}} - \left(A + \left[ \frac{\tanh(\alpha)}{\alpha} - 1 \right] \frac{1}{1-\mu^2}(A - \mu H)\right) = \Lambda_2. \nonumber
\end{align}
Notice that $\Lambda_1^\prime$ takes on the traditional Abelian form, since $[\Lambda_1^\prime, \pi_{\bm \lambda}^{(1)}] = 0$. Analogously, we obtain the following relevant operators:
 \begin{align}
 &Q_1^{\prime (1)} - \tilde{Q}_1^{\prime (1)} =  \sin^2(g\tau)(H^{(1)} - \mu A^{(1)}) - \sin^2(g\tau)(H^{(2)} - \mu A^{(2)}) \nonumber \\ &~- \cos(g \tau) \sin(g \tau) \omega [a_1^\dagger a_2 + a_1 a_2^\dagger + \mu(a_1 a_2 + a_1^\dagger a_2^\dagger)], \\
 &Q_2^{\prime (1)} - \tilde{Q}_2^{\prime (1)} =  \sin^2(g\tau) A^{(1)} \nonumber \\ &~- \sin^2(g\tau) A^{(2)} - \cos(g \tau) \sin(g \tau) \omega(a_1 a_2 + a_1^\dagger a_2^\dagger).
\end{align}
Taking the trace over subsystem 2 in the above expression and introcing them in Eq.~\eqref{onsagerSLD} together with the SLDs~\eqref{eq:lambdas1p}, we obtain the set of coefficients reported in Eq.~\eqref{eq:thermocoeffI} of Sec.~\ref{sec:effects}.

\bibliography{library2}

\begin{thebibliography}{90}
\bibitem [{not({\natexlab{a}})}]{noteH}%
  \BibitemOpen
  \href@noop {} {} ({\natexlab{a}}),\ \bibinfo {note} {{This unitary operation
  represents the interaction between reservoirs e.g. through an interface with
  Hamiltonian $H_\mathrm{int}$, such that $U \simeq e^{-i H_\mathrm{int}
  \tau}$, where $\tau$ is the collision time.}}\BibitemShut {Stop}%
\bibitem [{not({\natexlab{b}})}]{noteJ}%
  \BibitemOpen
  \href@noop {} {} ({\natexlab{b}}),\ \bibinfo {note} {{For example a bipartite
  interface consisting of two two-level systems weakly interacting thought a
  hopping Hamiltonian. Since the two-level systems are unnable to accomodate
  second-order coherence in their density operators, the squeezing current
  would be effectively blocked.}}\BibitemShut {Stop}%
\end{thebibliography}

\end{document}